\documentclass[pdftex,twocolumn,epjc3]{svjour3}          

\usepackage{amsmath}

\RequirePackage[T1]{fontenc}

\smartqed  

\RequirePackage{graphicx}
\RequirePackage{mathptmx}      
\RequirePackage{flushend}
\RequirePackage[numbers,sort&compress]{natbib}
\RequirePackage[colorlinks,citecolor=blue,urlcolor=blue,linkcolor=blue]{hyperref}

\journalname{Eur. Phys. J. C}

\begin{document}

\title{Tsallis statistics approach to the transverse momentum distributions \\ in p-p collisions}


\author{Maciej Rybczy\'nski\thanksref{e1,addr1}
        \and
        Zbigniew W\l odarczyk\thanksref{e2,addr1} 
}

\thankstext{e1}{e-mail: Maciej.Rybczynski@ujk.edu.pl}
\thankstext{e2}{e-mail: Zbigniew.Wlodarczyk@ujk.edu.pl}

\institute{Institute of Physics, Jan Kochanowski University, ul. Swietokrzyska 15, PL-25406~Kielce, Poland \label{addr1}      
}

\date{Received: date / Accepted: date}

\maketitle

\begin{abstract}
Transverse momentum distributions of negatively charged pions produced in p-p interactions at beam momenta 20, 31, 40, 80 and 158 GeV$/c$ are studied using the Tsallis distribution as a parametrization. Results are compared with higher energies data and changes of parameters with energy are determined. Different Tsallis-like distributions are compared. 
\end{abstract}

\section{Introduction}
\label{Sec:Intro}

Transverse momentum ($p_T$) distributions of identified hadrons are the most common tools used to study the dynamics of high energy collisions. The p-p interactions are used as a baseline and are important to understand the particle production mechanism~\cite{Becattini:1997rv}. In the framework of Tsallis statistics~\cite{Tsallis:1987eu,Cleymans:2011in,Cleymans:2012ya} the momentum distribution is given by

\begin{align}
\frac{d^{3}N}{dp^{3}} & = \frac{gV}{\left(2\pi\right)^{3}}\left[1+\left(q-1\right)\frac{E-\mu}{T}\right]^{\frac{q}{1-q}}\xrightarrow{q\rightarrow 1} \nonumber \\
                      & \frac{gV}{\left(2\pi\right)^{3}}\exp\left(-\frac{E-\mu}{T}\right),
\label{eq:ptdist}
\end{align}
where $T$ and $\mu$ are the temperature and the chemical potential, $V$ is the volume and $g$ is the degeneracy factor. In this form, Eq.~(\ref{eq:ptdist}) is usually supposed to represent a nonextensive generalization of the
Boltzmann-Gibbs exponential distribution, $\exp\left(-E/T\right)$, with $q$ being a new parameter, in addition to previous "temperature" $T$. Such an approach is known as nonextensive statistics~\cite{Tsallis:1987eu} in which the parameter $q$ summarily describes all features causing a departure from the usual Boltzmann-Gibbs statistics. In particular it was shown in~\cite{Wilk:1999dr} that $q-1=\textrm{Var}\left(T\right)/\langle T\rangle^{2}$ and directly describes intrinsic fluctuations of temperature (however, the Tsallis distribution also emerges from a number of other more dynamical mechanisms, for example see~\cite{Wilk:2012zn} for more details and references).
This approach has been shown to be very successful in describing  multiparticle production processes of a different kind (see~\cite{Wilk:2012zn,Wilk:2008ue} for recent reviews).
In terms of transverse momentum, transverse mass, $m_{T}=\sqrt{m^{2}+p_{T}^{2}}$, and rapidity $y$, Eq.~(\ref{eq:ptdist}) becomes

\begin{align}
\frac{d^{2}N}{p_{T}dp_{T}dy} & = gV\frac{m_{T}\cosh\left(y\right)}{\left(2\pi\right)^{2}}\times \nonumber \\  
                  & \times\left[1+\left(q-1\right)\frac{m_{T}\cosh\left(y\right)-\mu}{T}\right]^{\frac{q}{1-q}}.
\label{eq:ptdist2}
\end{align}

It has been shown repeatedly that the Tsallis distribution gives an excellent description of $p_{T}$ spectra measured in p-p collisions at RHIC ($\sqrt{s} = 62.4$ and $200$~GeV) and LHC ($\sqrt{s} = 0.9$, $2.76$ and $7.0$~TeV) energies~\cite{Cleymans:2011in, Sena:2012ds, Adare:2010fe, Khandai:2013gva, Wong:2012zr}. In particular changes in the transverse momentum distribution with energy (used data at energies $0.54$, $0.9$, $2.36$ and $7$ TeV) are studied using the Tsallis distribution~(\ref{eq:ptdist2}) as a parametrization~\cite{Cleymans:2013rfq}. In this paper we extend this analysis to transverse momentum spectra obtained in p-p collisions at  $\sqrt{s} = 6.27$, $7.74$, $8.76$, $12.32$ and $17.27$~GeV by the NA61/SHINE collaboration~\cite{Abgrall:2013qoa}~\footnote{Recently, the experimental results on inclusive spectra of negatively charged pions produced in inelastic p-p interactions at beam momenta $20$, $31$, $40$, $80$ and $158$~GeV$/c$ were presented~\cite{Abgrall:2013qoa}. The measurements were performed using the large acceptance NA61/SHINE hadron spectrometer at the CERN Super Proton Synchrotron.  Numerical results corresponding to the two dimensional spectra in transverse momentum and rapidity corrected for experimental biases were given in Ref.~\cite{na61data}.}. In addition to possibility of study collisions at low incident energies, the measurements performed by NA61/SHINE collaboration allow us to study the low-$p_{T}$ part of spectra. The values of $T$ and $V$ are very sensitive to the low-$p_{T}$ part of the transverse momentum distribution and extending the analysis to lower $p_{T}$ could bring much  clarification here.

\section{Analysis of transverse momentum distributions}
\label{Sec:Analysis}

Transverse momentum spectra of negatively charged pions are fitted using Tsallis distribution given by Eq.~(\ref{eq:ptdist2}) with $g_{\pi^{-}}=1$ and $\mu=0$. It is worth to be noted that the variable $T$ and $V$ are  functions of $\mu$ at fixed values of $q$,

\begin{equation}
T=T_{0}+\left(q-1\right)\mu,
\label{eq:EqT}
\end{equation}

\begin{equation}
V=V_{0}\left[1+\left(q-1\right)\mu/T_{0}\right]^{q/\left(1-q\right)}=V_{0}\left(T/T_{0}\right)^{q/\left(1-q\right)}
\label{eq:EqV}
\end{equation}
and they can be calculated if the parameters $T=T_{0}$ and $V=V_{0}$ at $\mu=0$ are known~\cite{Cleymans:2013rfq}.

\begin{figure} %
\begin{minipage}{\columnwidth}
\centering
\centerline{\includegraphics[width=0.97\columnwidth]{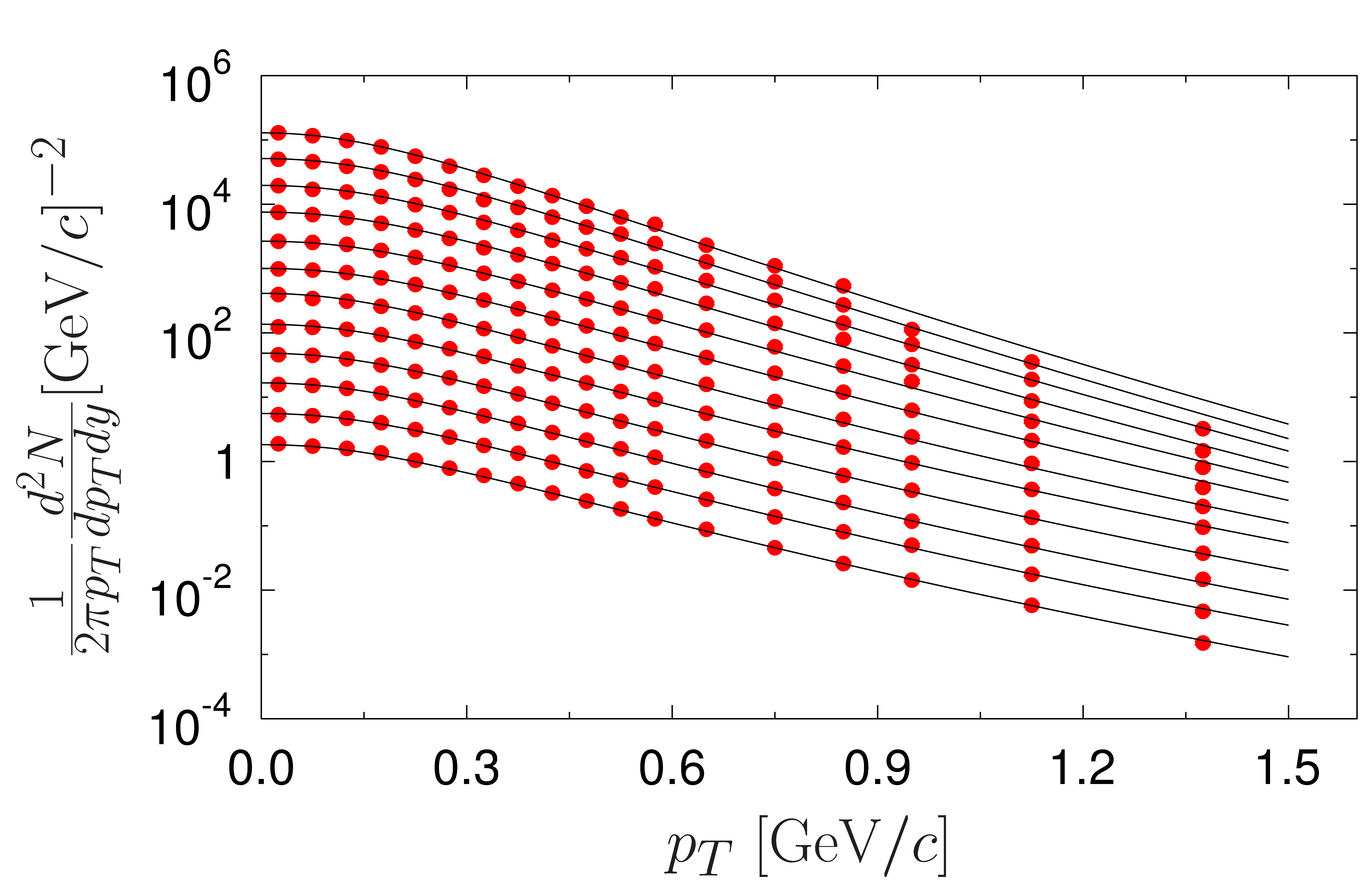}} 
\end{minipage}
\caption{\small (Color online) Transverse momentum distributions of negatively charged pions produced in p-p collisions as obtained by the NA61/SHINE collaboration~\cite{Abgrall:2013qoa} at $\sqrt{s} = 17.27$~GeV in rapidity intervals $0.2k < y < 0.2\left(k+1\right)$ where $k = 0,\cdots, 11$ from the bottom. Data points for different rapidity bins were scaled by $3^{k}$ for better readability.}
\label{fig:fig_158_v3}
\end{figure}

\begin{figure} %
\begin{minipage}{\columnwidth}
\centering
\centerline{\includegraphics[width=0.97\columnwidth]{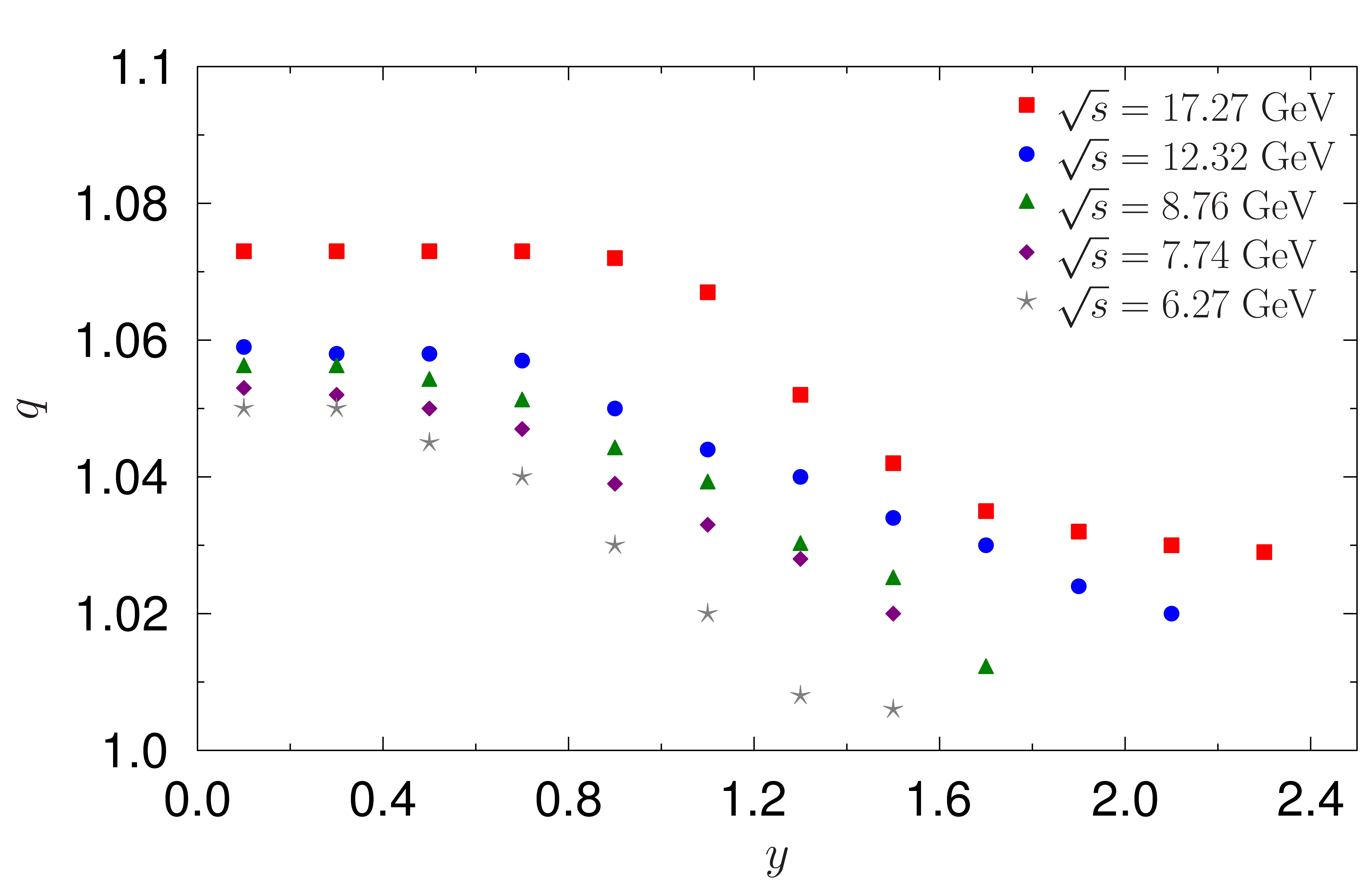}} 
\end{minipage}
\caption{\small (Color online) The values of the nonextensivity parameter $q$, as a function of rapidity obtained from fits to the transverse momentum distributions at different energies.}
\label{fig:q_y}
\end{figure}

\begin{figure} %
\begin{minipage}{\columnwidth}
\centering
\centerline{\includegraphics[width=0.97\columnwidth]{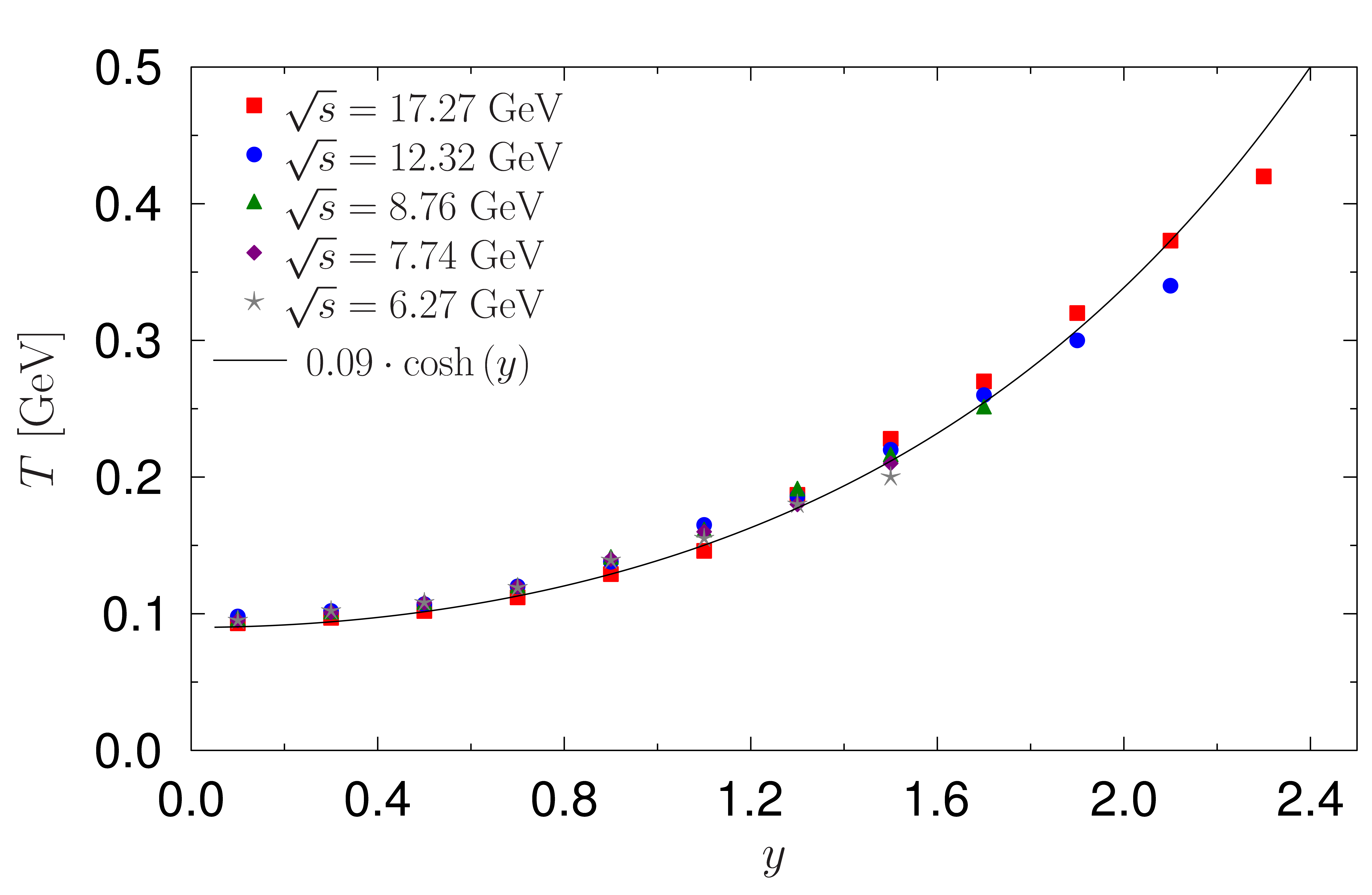}} 
\end{minipage}
\caption{\small (Color online) The values of the temperature parameter, $T$, as a function of rapidity obtained
from fits to the transverse momentum distributions at different energies.}
\label{fig:T_y}
\end{figure}

The Tsallis distribution describes the transverse momentum distributions of negatively charged pions in p-p collisions as obtained by the NA61/SHINE collaboration~\cite{Abgrall:2013qoa} in all rapidity intervals remarkably well as shown in Fig.~\ref{fig:fig_158_v3}. The values of nonextensivity parameters $q$ needed to describe the transverse momentum distributions of negatively charged pions are shown in Fig.~\ref{fig:q_y}. The values of temperature parameter $T$ for different energies and rapidity intervals are shown in Fig.~\ref{fig:T_y}.
The temperature parameter $T$ shows a clear rapidity dependence which we have parametrized as $T\simeq 0.09\cosh\left(y\right)$.

\section{Energy dependence of parameters}
\label{Sec:EnDep}

\begin{figure} %
\begin{minipage}{\columnwidth}
\centering
\centerline{\includegraphics[width=0.97\columnwidth]{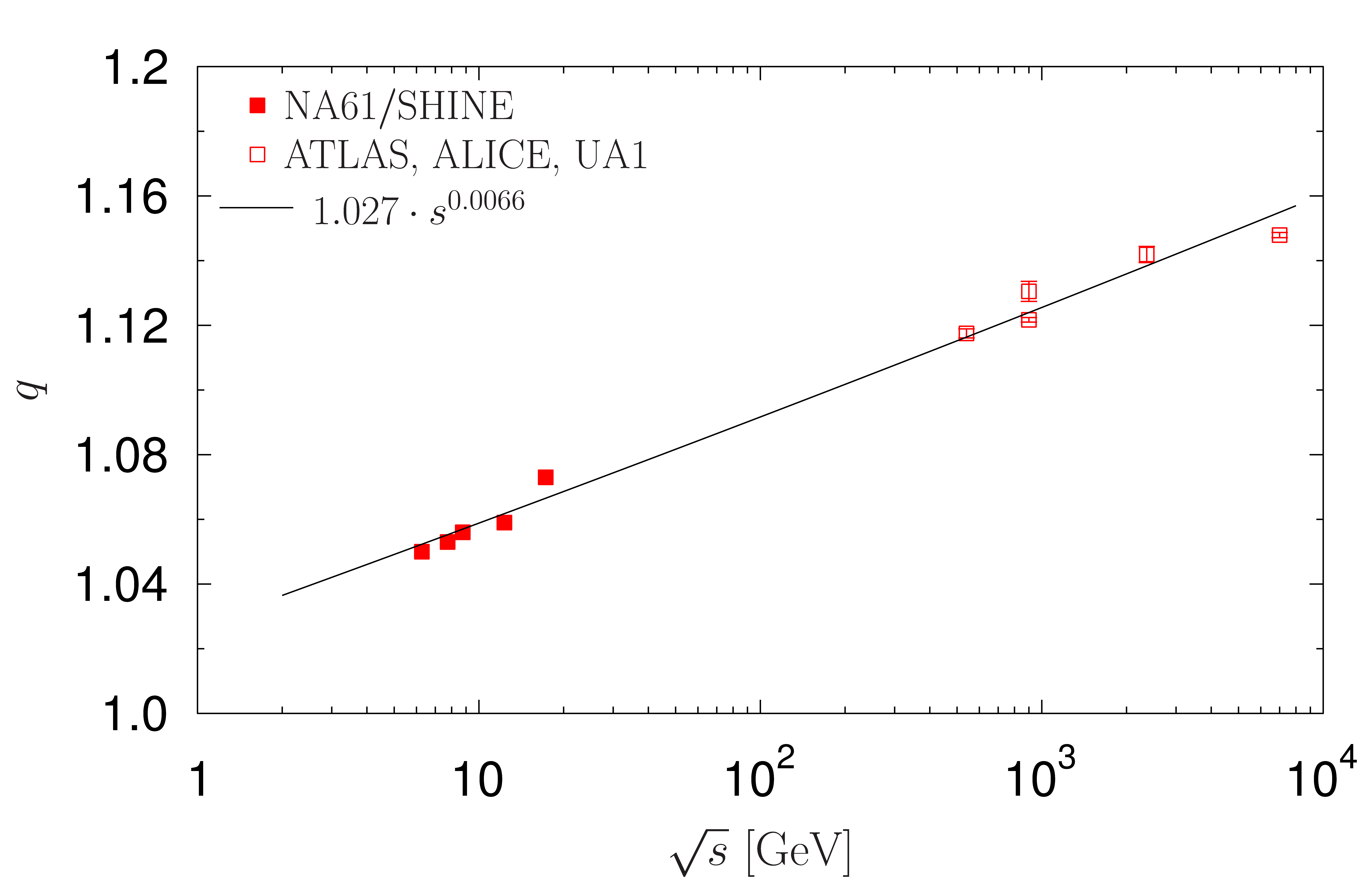}} 
\end{minipage}
\caption{\small (Color online) Energy dependence of the parameter $q$ appearing in the Tsallis distribution. Open points are from ATLAS, ALICE and UA1 Collaborations data (taken from Ref.~\cite{Cleymans:2013rfq}). Solid points are from NA61/SHINE Collaboration data~\cite{Abgrall:2013qoa}. Data are fitted by Eq.~(\ref{eq:qs}).}
\label{fig:q_ecm}
\end{figure}

\begin{figure} %
\begin{minipage}{\columnwidth}
\centering
\centerline{\includegraphics[width=0.97\columnwidth]{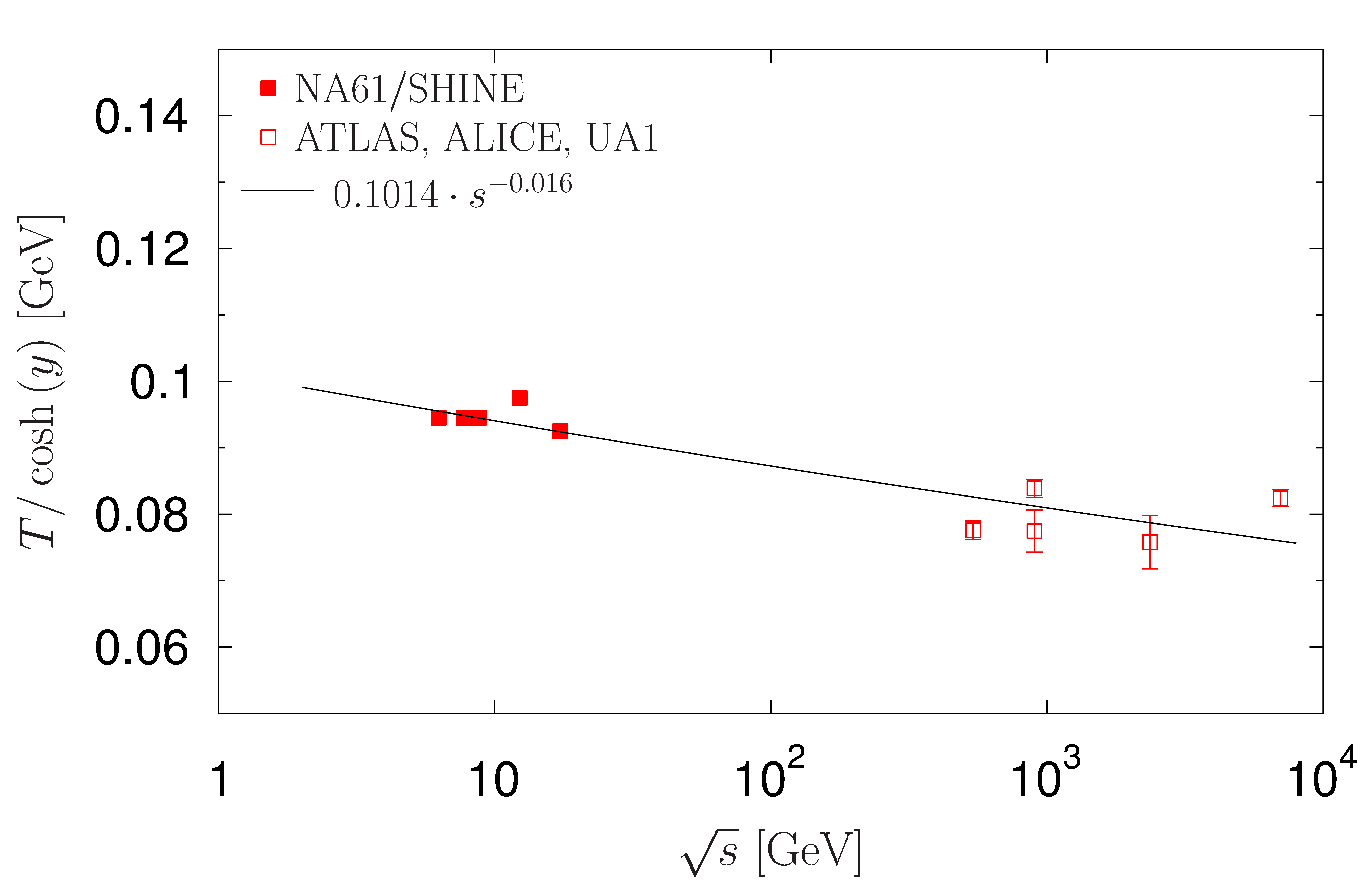}} 
\end{minipage}
\caption{\small (Color online) Energy dependence of the temperature parameter $T$ appearing in the Tsallis distribution. Open points are from ATLAS, ALICE and UA1 Collaborations data (taken from Ref.~\cite{Cleymans:2013rfq}). Solid points are from NA61/SHINE Collaboration data~\cite{Abgrall:2013qoa}. Data are fitted by Eq.~(\ref{eq:Ts}).}
\label{fig:T1_ecm}
\end{figure}

\begin{figure} %
\begin{minipage}{\columnwidth}
\centering
\centerline{\includegraphics[width=0.97\columnwidth]{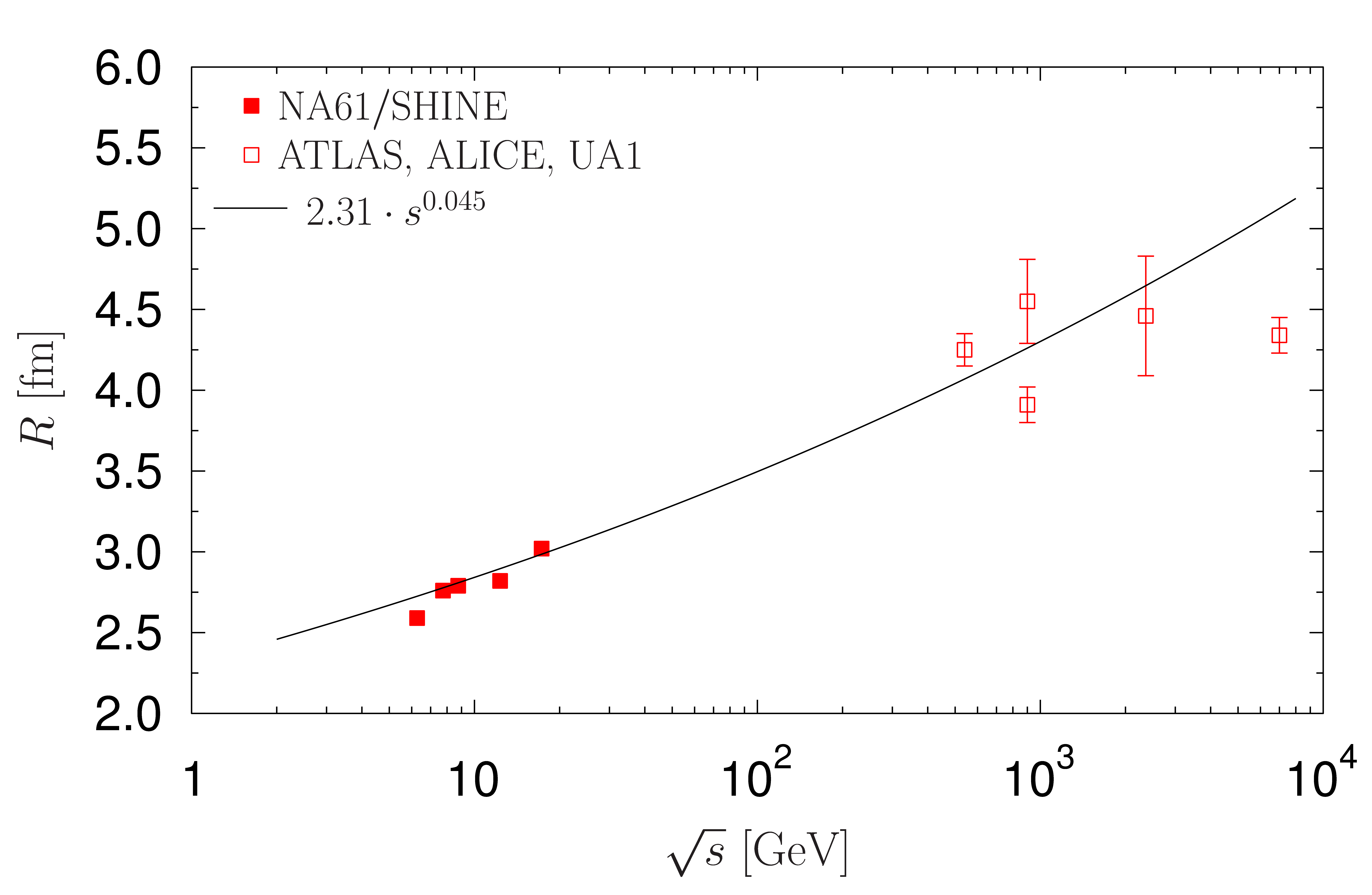}} 
\end{minipage}
\caption{\small (Color online) Energy dependence of the radius $R$ appearing in the volume factor, $V=4/3\pi R^{3}$. Open points are from ATLAS, ALICE and UA1 Collaborations data (taken from Ref.~\cite{Cleymans:2013rfq}). Solid points are from NA61/SHINE Collaboration data~\cite{Abgrall:2013qoa}. Data are fitted by Eq.~(\ref{eq:Rs}).}
\label{fig:r_ecm}
\end{figure}

The energy dependence of the various parameters is displayed in Figs.~\ref{fig:q_ecm}, \ref{fig:T1_ecm} and \ref{fig:r_ecm}. For comparison with higher energy data~\cite{Cleymans:2013rfq} which are for mid-rapidity $y=0$, we show parameters as evaluated for rapidity interval $0<y<0.2$. All analysed parameters show a clear but weak energy dependence which we have parametrized as 

\begin{equation}
q\left(s\right)=1.027\left(\sqrt{s}\right)^{0.01326}
\label{eq:qs}
\end{equation}

\begin{equation}
T\left(s\right)=0.1014\left(\sqrt{s}\right)^{-0.03262}
\label{eq:Ts}
\end{equation}

\begin{equation}
R\left(s\right)=\left(\frac{3V\left(s\right)}{4\pi}\right)^{1/3} =2.31\left(\sqrt{s}\right)^{0.09}
\label{eq:Rs}
\end{equation}

The value of $R$ is not necessarily related to the size of the system as deduced from a HBT analysis~\cite{Aamodt:2010jj,Aggarwal:2010aa} but serves to fix the normalization of the distribution (\ref{eq:ptdist2}). In particular, we have

\begin{align}
\frac{dN}{dy}\Biggl\vert_{y=0} & =\frac{gVT}{\left(2\pi\right)^{2}}\left[1+\left(q-1\right)\frac{m}{T}\right]^{\frac{1}{1-q}}\times \nonumber \\
& \times\frac{\left(2-q\right)m^{2}+2mT+2T^{2}}{\left(2-q\right)\left(3-2q\right)}.
\label{eq:dNdy}
\end{align}

For evaluated above energy dependence of parameters $q\left(s\right)$, $T\left(s\right)$ and $R\left(s\right)$ given by Eqs.~(\ref{eq:qs}-\ref{eq:Rs}) we have 

\begin{equation}
\frac{dN}{dy}\Biggl\vert_{y=0}\simeq 0.1+0.56\left(\sqrt{s}\right)^{0.24}.
\label{eq:dNdy2}
\end{equation}

\begin{figure} %
\begin{minipage}{\columnwidth}
\centering
\centerline{\includegraphics[width=0.97\columnwidth]{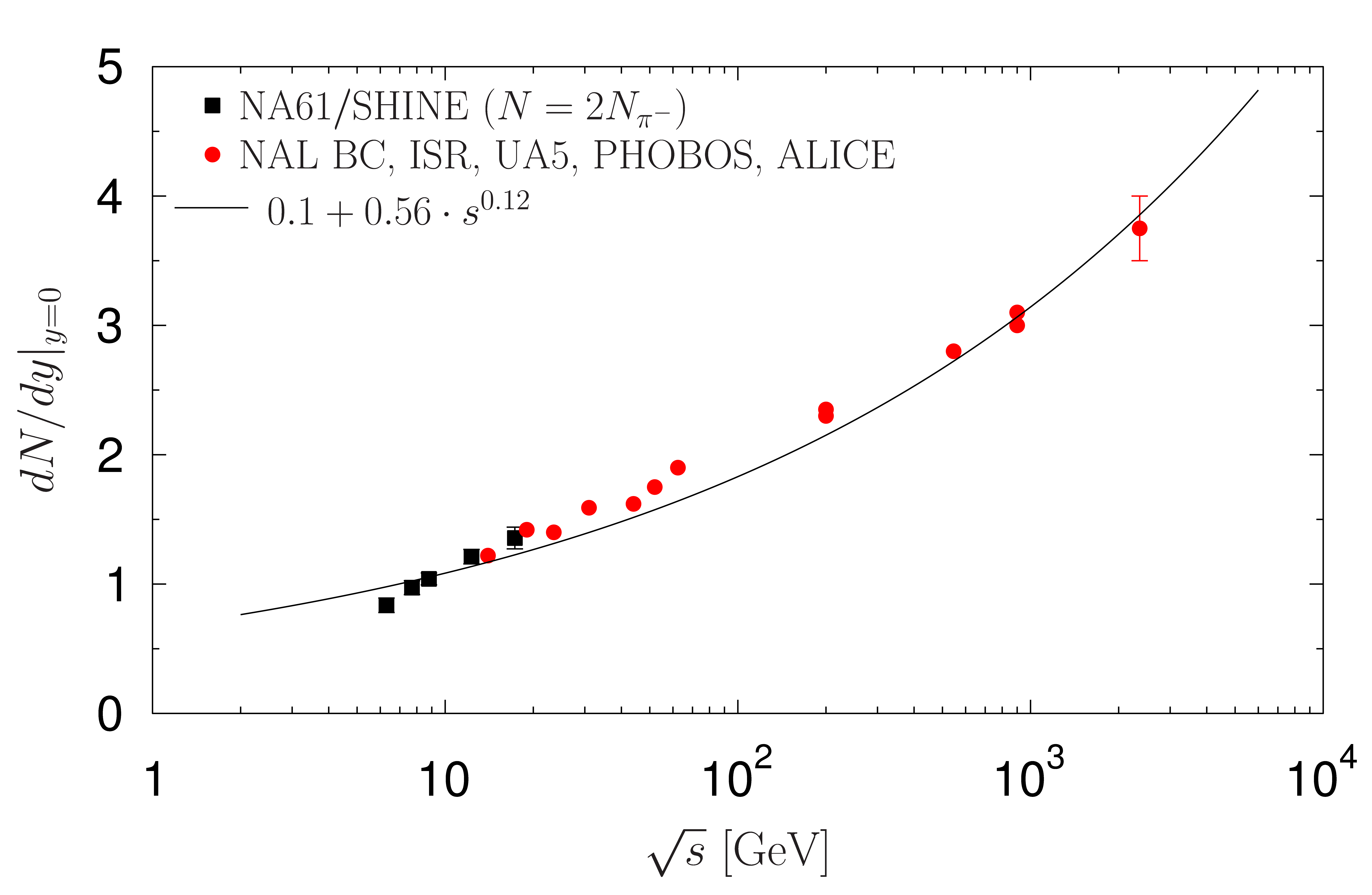}} 
\end{minipage}
\caption{\small (Color online) $dN/dy$ of charged particles produced in the central rapidity region as a function of center-of mass energy in $p-p$ and $p-\bar{p}$ collisions. Energy dependence given by Eq.~(\ref{eq:dNdy2}) is compared with inelastic measurements  from NA61/SHINE~\cite{Abgrall:2013qoa} ($p-p$),  NAL Bubble Chamber ($p-\bar{p}$), ISR ($p-p$), UA5 ($p-\bar{p}$), PHOBOS ($p-p$) and ALICE ($p-p$) experiments taken from compilation~\cite{Khachatryan:2010us}.}
\label{fig:dndy}
\end{figure}

Energy dependence of $dN/dy$ in the central rapidity region in comparison with inelastic measurements is shown in Fig.~\ref{fig:dndy}.

\begin{figure} %
\begin{minipage}{\columnwidth}
\centering
\centerline{\includegraphics[width=0.97\columnwidth]{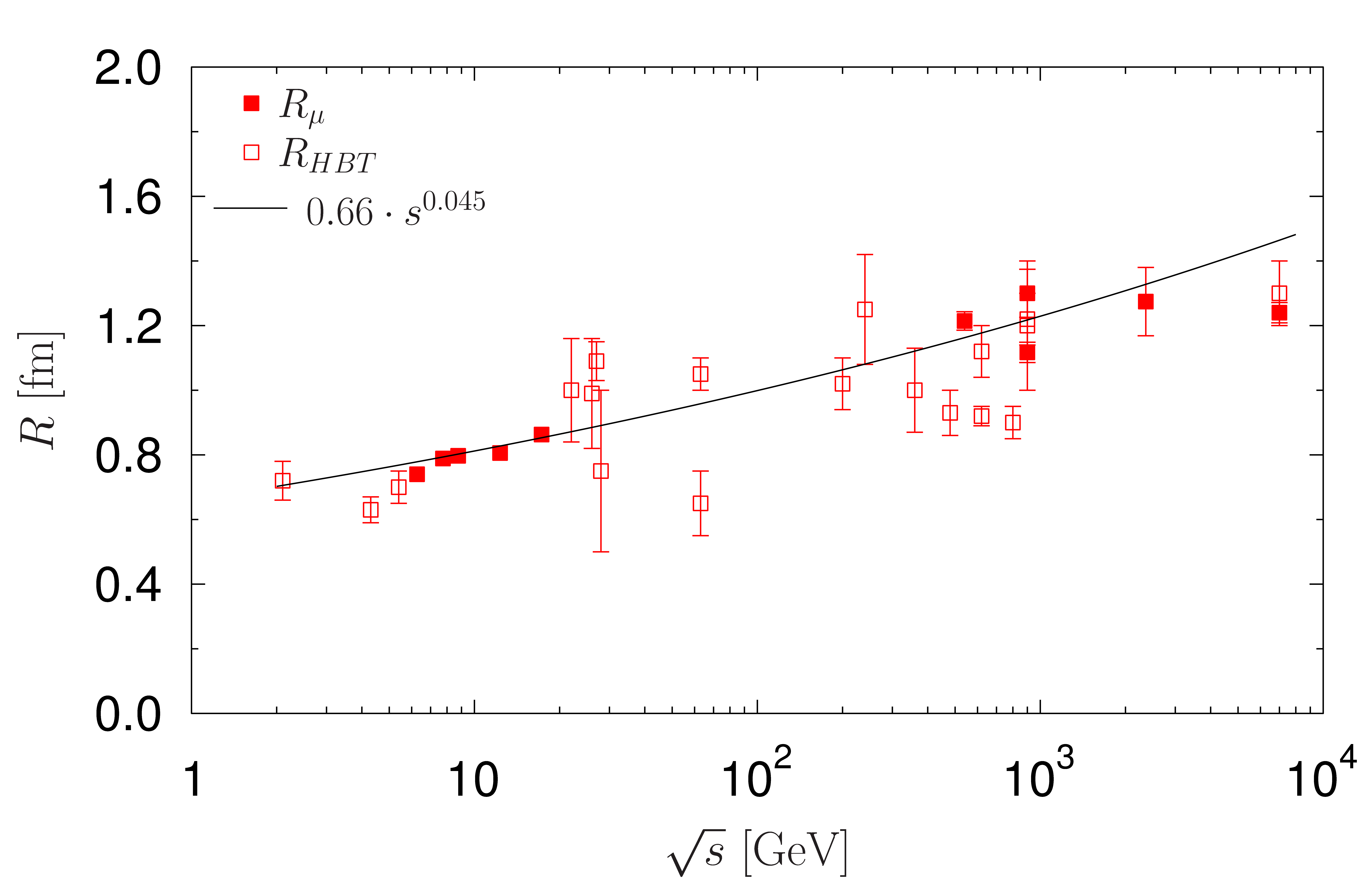}} 
\end{minipage}
\caption{\small (Color online) Energy dependence of the radius $R_{\mu}=R_{\mu=0}/3.5$ (solid points) in comparison with HBT measurements of source radii obtained in hadron-hadron reactions~\cite{kageya} (open points).}
\label{fig:r_hbt}
\end{figure}

We can treat the size of the system, $R$, more seriously. The radius given by Eq.~(\ref{eq:Rs}) is calculated for $\mu=0$. For other values of chemical potential, the size is smaller (cf. Eqs.~(\ref{eq:EqT}) and~(\ref{eq:EqV})). Comparing $R\left(s\right)$ with experimental data deduced from HBT analysis we can see that $R_{HBT}\simeq R/\kappa$, where $\kappa=3.5$. In Fig.~\ref{fig:r_hbt} we displayed $R\left(s\right)/\kappa$ in comparison with data obtained from HBT analysis~\cite{kageya}.

Following this observation we assume

\begin{equation}
V_{\mu=0}=V_{\mu}\cdot\kappa^{3}
\label{eq:Vmu}
\end{equation}
and from Eqs.~(\ref{eq:EqT}) and~(\ref{eq:EqV}) we have

\begin{equation}
\mu=\frac{T_{\mu=0}}{q-1}\left(\kappa^{3\left(q-1\right)/q}-1\right)
\label{eq:mukappa}
\end{equation}
and using parametrizations (\ref{eq:qs}) and (\ref{eq:Ts}) we have energy dependence of chemical potential in the form

\begin{equation}
\mu\left(s\right)\simeq 0.39\left(\sqrt{s}\right)^{-0.022}.
\label{eq:mus}
\end{equation}

\section{Different parametrizations}
\label{Sec:DifPar}

Almost fifty years ago Hagedorn develop a statistical description of momentum spectra observed in multiparticle production processes~\cite{Hagedorn}. Hagedorn's approach predicts an exponential decay of momentum distribution 
\begin{equation}
E\frac{d^{3}N}{dp^{3}}\simeq C\exp\left(-\frac{p_{T}}{T}\right)
\label{eq:hag1}
\end{equation}
for transverse momenta, whereas in experiments one observes non-exponential behavior for large transverse momenta. Subsequently, Hagedorn proposed  the  "QCD inspired" empirical formula describing the data of the invariant momentum distribution of hadrons as a function of $p_{T}$ over a wide range~\cite{Hagedorn:1983wk}:
\begin{equation}
E\frac{d^{3}N}{dp^{3}} = C\left(1+\frac{p_{T}}{p_{0}}\right)^{-n}\rightarrow 
\begin{cases}
\exp\left(-np_{T}/p_{0}\right) & \text{ for } p_{T}\rightarrow 0 \\ 
\left(p_{T}/p_{0}\right)^{-n} & \text{ for } p_{T}\rightarrow \infty
\end{cases}
\label{eq:hag2}
\end{equation}
with  $C$, $p_{0}$ and $n$ being fit parameters. This becomes pure exponential for small $p_{T}$ and pure power law for large $p_{T}$. For $n=q/\left(q-1\right)$  and $p_{0}=T/\left(q-1\right)$, the Hagedorn formula (\ref{eq:hag2}) coincides with Tsallis distribution~\cite{Tsallis:1987eu},
\begin{equation}
E\frac{d^{3}N}{dp^{3}} = C\left[1-\left(1-q\right)\frac{p_{T}}{T}\right]^{\frac{q}{1-q}}.
\label{eq:tsall}
\end{equation}

The  basic conceptual difference between (\ref{eq:hag2}) and (\ref{eq:tsall}) is in the underlying physical picture. In (\ref{eq:hag2}) the low-$p_{T}$ region is controlled by soft physics represented by some unknown 
unperturbative theory or model, and the high-$p_{T}$ region is governed by hard physics represented
by perturbative QCD. In (\ref{eq:tsall}), the nonextensive formula works in the whole range of $p_{T}$ and it is not derived from some particular theory. It is only a generalization of the regular statistical mechanics and just offers the kind of universal unifying principle, namely the existence of some kind of equilibrium affecting all scales of $p_{T}$, which is described by two parameters, $T$ and $q$. The temperature $T$ characterize its mean properties and the parameter $q$, known as the nonextensivity parameter, expresses action of the potentially non-trivial long range effects believed to be caused by fluctuations~\cite{Wilk:1999dr} (but also by some correlations or long memory effects~\cite{Tsallis:1987eu}). It is worth to be noted that the invariant momentum distribution in the form (cf. Eq.(\ref{eq:ptdist}))

\begin{equation}
E\frac{d^{3}N}{dp^{3}}=\frac{gV}{\left(2\pi\right)^{3}}\left[1+\left(q-1\right)\frac{E}{T}\right]^{\frac{q}{1-q}},
\label{eq:ptdist3}
\end{equation}
result in  Eq.~(\ref{eq:ptdist2}) without pre-factor $m_{T}\cosh\left(y\right)$ in the right hand side of the equation. For the non-relativistic energies ($E=p^{2}/\left(2m\right)$), Eq.~(\ref{eq:ptdist3}) corresponds to Tsallis distribution

\begin{equation}
E\frac{d^{3}N}{dp^{3}}=\frac{gV}{\left(2\pi\right)^{3}}\left[1+\left(q-1\right)\frac{p^{2}}{2mT}\right]^{\frac{q}{1-q}},
\label{eq:ptdist4}
\end{equation}
originated from multiplicative noise~\cite{Biro:2004qg,Anteneodo}~\footnote{The Langevin equation $dp/dt + \gamma\left(t\right)p=\xi\left(t\right)$ where both $\gamma\left(t\right)$ and $\xi\left(t\right)$ denote stochastic processes (traditional multiplicative noise and additive noise, respectively) leads to a power-law tail of the distribution for sufficiently large momenta. As shown in~\cite{Biro:2004qg} in the case of $\textrm{Cov}\left(\gamma,\xi\right)=0$ and $E\left(\xi\right)=0$ (i.e., for, respectively, no correlation between noises and no drift term due to the additive noise) the solution is given by the non-normalized Tsallis distribution for the variable $p^{2}$.}.

Exponential function Eq.~(\ref{eq:hag1}) described data only in the limited range of transverse momentum, $0.15<p_{T}<0.6$~\cite{Abgrall:2013qoa}. As shown in Fig.~\ref{fig:fig_158_v3}, the Tsallis distribution given by Eq.~(\ref{eq:ptdist2}) describes all $p_T$ range remarkably well. 

All Tsallis-like distributions lead to a power law tail

\begin{equation}
\frac{d^{2}N}{p_{T}dp_{T}dy}\propto p_{T}^{-n}
\label{eq:pow}
\end{equation}
of the distribution for sufficiently large transverse momenta. The difference between them can be seen in low $p_{T}$ region, where

\begin{equation}
\frac{d^{2}N}{p_{T}dp_{T}dy}\propto
\begin{cases}
\alpha-\beta p_{T} +\gamma p_{T}^{2} & \text{ for Eqs.~(\ref{eq:hag1}), (\ref{eq:hag2}), (\ref{eq:tsall})}  \\ 
\alpha-\gamma p_{T}^{2} & \text{ for Eqs.~(\ref{eq:ptdist}), (\ref{eq:ptdist3}), (\ref{eq:ptdist4})} 
\end{cases}
\label{eq:pow2}
\end{equation}

\begin{figure} %
\begin{minipage}{\columnwidth}
\centering
\centerline{\includegraphics[width=0.97\columnwidth]{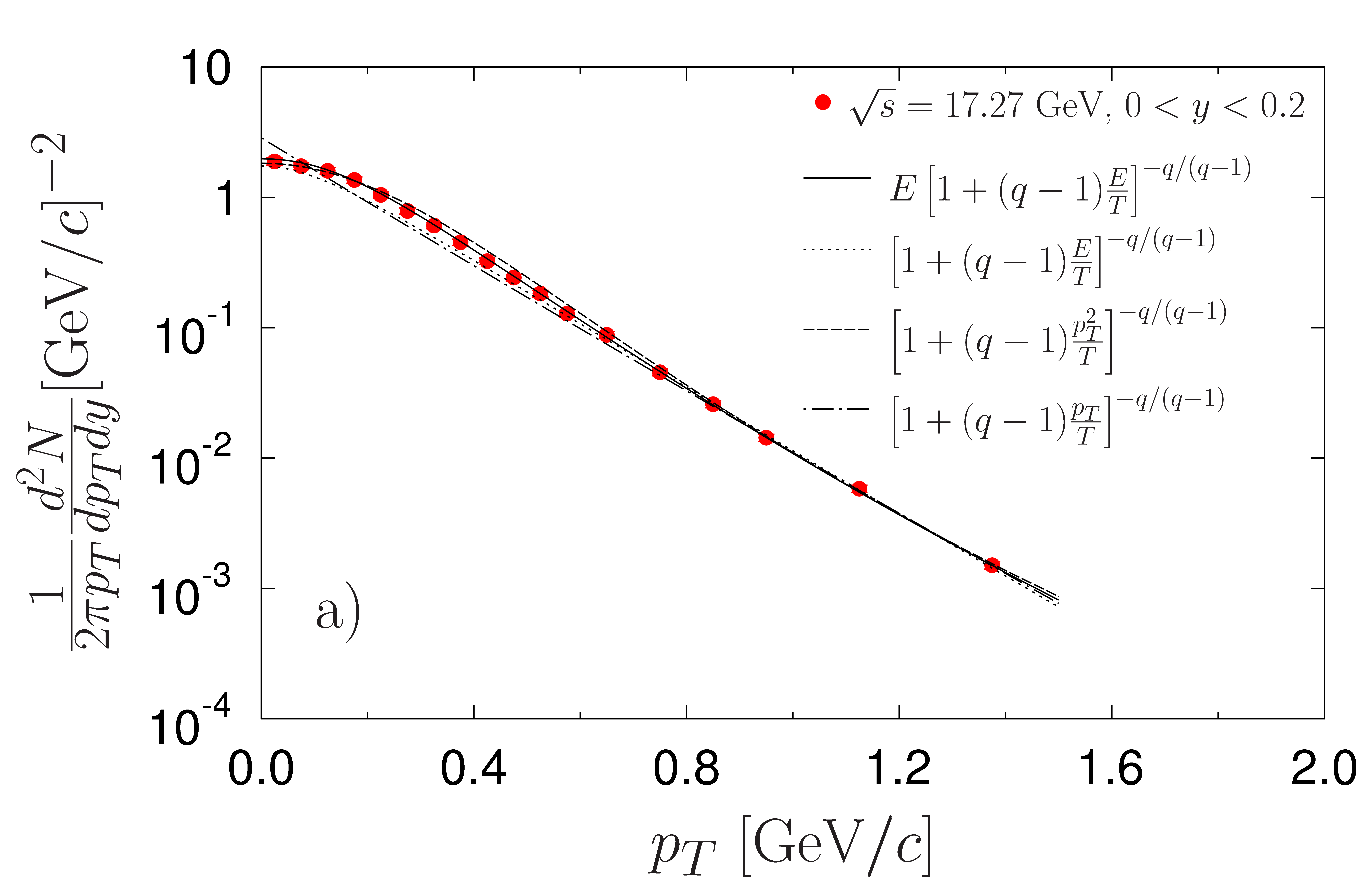}} 
\centerline{\includegraphics[width=0.97\columnwidth]{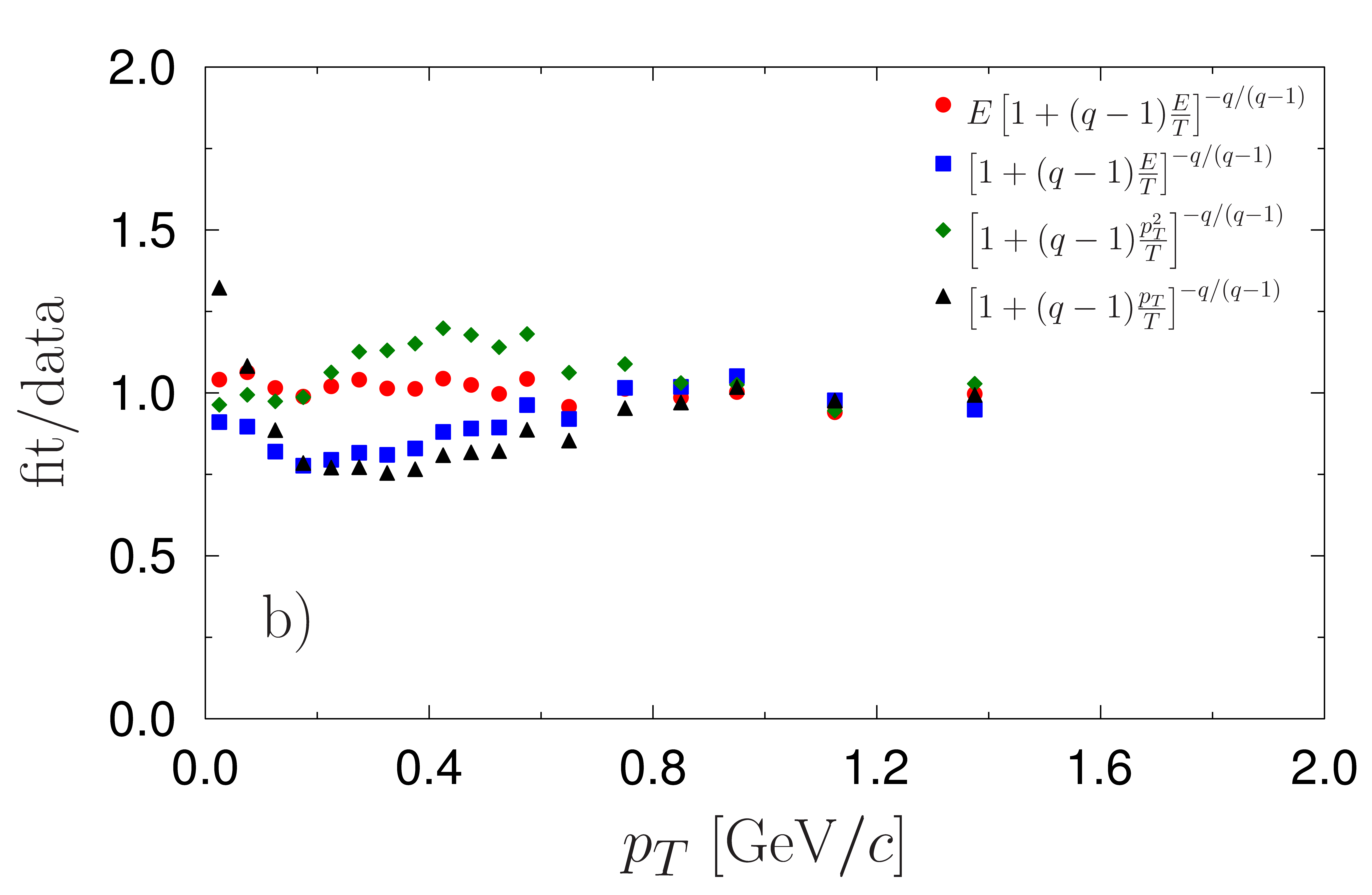}}
\end{minipage}
\caption{\small (Color online) Panel (a) - transverse momentum distribution of negatively charged pions produced in $p-p$ collisions at $\sqrt{s}=17.27$~GeV in the rapidity interval $0<y<0.2$~\cite{Abgrall:2013qoa} fitted by different parametrizations (with normalization at hight $p_{T}$ region). Panel  (b) - ratio $fit/data$ for the results presented in panel (a).}
\label{fig:fig_158_tyl}
\end{figure}

Parameters $\alpha$, $\beta$ and $\gamma$ are positive valued functions of $q$ and $T$ (in case of Eq.(\ref{eq:ptdist}), $T<m$ is required for $\gamma>0$ ).    
In low $p_{T}$ region, Tsallis-like distributions with variable $p_{T}^{2}$ differs from the one expressed in variable $p_{T}$. Comparison of  different parametrizations is shown in Fig.~\ref{fig:fig_158_tyl}.

\section{Discussion and conclusions}
\label{Sec:DisCon}

In conclusion, the Tsallis distribution, Eq.~(\ref{eq:ptdist2}) leads to an excellent description of data on transverse momentum. By comparing results from NA61/SHINE~\cite{Abgrall:2013qoa} to the results obtained at higher energies~\cite{Cleymans:2013rfq} it has been possible to extract energy dependence of the parameters $q$, $T$ and $R$. A consistent picture emerges from comparison of fits using the Tsallis distribution in wide range of energies.

Different parametrizations lead not only to different quality of fits but also to different values of parameters. In Ref.~\cite{Abgrall:2013qoa} experimental data are fitted by exponential distribution~(\ref{eq:hag1}) in limited range of transverse momenta ($0.15<p_{T}<0.6$~GeV$/c$) evaluating temperature parameters seemingly larger than our estimate based on parametrization (\ref{eq:ptdist2}). Such difference in values of temperature parameters is fully understandable. For distributions with the same mean transverse momentum, $\langle p_{T}\rangle$, the parameter $T_{exp}$ evaluated from  Eq.~(\ref{eq:hag1}) is connected with parameter $T$ evaluated from Eq.~(\ref{eq:ptdist2}) by the relation

\begin{equation}
T_{exp}\simeq a+b\cdot T,
\end{equation}
where, numerically, $a=0.31-0.654q+0.354q^{2}$ and $b=27.35-55q+29.07q^{2}$. Moreover, it is remarkable to notice that parametrization (\ref{eq:ptdist}) proposed by Cleymans~\cite{Cleymans:2011in, Cleymans:2012ya} is for momentum distribution, $d^{3}N/dp^{3}$ while the other Tsallis-like parametrizations~(\ref{eq:hag2})-(\ref{eq:ptdist4}) are for invariant distribution $E d^{3}N/dp^{3}$.

\begin{acknowledgements}
This research was supported by the National Science Center (NCN) under contracts: 2011/03/B/ST2/02617 \newline and 2012/04/M/ST2/00816.
\end{acknowledgements}

\end{document}